\documentclass[seceq]{ptptex}
\usepackage{wrapft}
\usepackage{graphicx}

\begin{document}
\begin{flushright}
{KOBE-TH-04-04}\\
\end{flushright}
\vspace{5mm}
\begin{center}
{\large \bf
Higgsless Gauge Symmetry Breaking with a Large Mass Hierarchy}
\vspace{10mm} \\
Tomoaki Nagasawa$^{1,}$\footnote{E-mail: nagasawa@phys.sci.kobe-u.ac.jp}
 and Makoto Sakamoto$^{2,}$\footnote{E-mail: sakamoto@phys.sci.kobe-u.ac.jp}
 \\
{\small \it
${}^1 $Graduate School of Science and Technology, Kobe University,
Rokkodai, Nada, \\ Kobe 657-8501, Japan
\\
${}^2 $Department of Physics, Kobe University,
Rokkodai, Nada, Kobe 657-8501, Japan
}
\vspace{4mm}
\\
\begin{minipage}[t]{120mm}
\baselineskip 2pt 
{\small
\quad \ 
We  propose a mechanism of Higgsless gauge symmetry breaking with a large mass hierarchy.
We consider a 5D gauge theory on an orbifold $S^1/Z_2$. 
The gauge symmetry is broken by orbifolding and also nontrivial boundary conditions at fixed points.
All 4D modes which survive at low energies are found to be localized around fixed points. 
Supersymmetry plays an important role in our mechanism.
The tree-level unitarity in our model is briefly discussed.
}
\end{minipage}
\end{center}
\vspace{4mm}
\section{Introduction}

New possibilities for symmetry breaking without Higgs fields have been proposed in the context of field theories in higher dimensions \cite{shark-schwarz, Hosotani, STT,phasestructure,BC,Murayama,GUTsOrbifold}.
Attractive models of Grand Unified Theories (GUTs) on orbifolds have been constructed, avoiding common problems of four-dimensional GUTs \cite{GUTsOrbifold}.
The GUT symmetry breaking scale is on the order of the inverse length of the compact space, but the origin of the weak scale is still a mystery.

A higher dimensional scenario to solve the hierarchy problem has been proposed by Randall and Sundrum (warped compactification).\cite{R-S}
In the scenario, all mass scales except for the scale of gravity are reduced to the weak scale by a warped factor, so that a large mass hierarchy in mass spectrum does not appear.
According to this scenario, various attempts have been made to construct realistic models.\cite{Nomura}

Another interesting possibility for the solution to the gauge hierarchy problem has been suggested by Shaposhnikov and Tinyakov.\cite{Shaposhnikov0102161}\footnote{We thank C. S. Lim for drawing our attention to the paper of Shaposhnikov and Tinyakov.}
\ They have considered a gauge invariant theory with a noncompact extra dimension and shown that massless modes disappear from the theory, while massive modes are organized in such a way that one is much lighter than the others.
The extension to GUTs is not, however, straightforward.

In this Letter, we push the idea of Shaposhnikov and Tinyakov \cite{Shaposhnikov0102161} forward, and study a possibility to solve the gauge hierarchy problem in a GUT  scenario, based on 5D gauge theories on orbifolds with nontrivial boundary conditions (BC's) at fixed points.
We present a toy model in which a large mass hierarchy of the spectrum naturally arises without a fine tuning.
All 4D modes which survive at low energies are localized around fixed points.
Other unwanted KK modes acquire masses on the order of a GUT scale and decouple at low energies, as desired.
In the next section, we examine one-dimensional $N=2$ supersymmetric quantum mechanics on a circle $S^1$ with two point singularities, and clarify allowed BC's compatible with supersymmetry. 
The mass spectrum for each BC is studied in some detail.
In Section 3, we demonstrate a toy model that possesses gauge symmetry breaking with a large mass hierarchy, with the help of the analysis of Section 2.
Section 4 is devoted to summary and discussions.

\section{Supersymmetric quantum mechanics on $S^1$ with point singularities}

In this section, we consider one-dimensional quantum mechanics on a circle $S^1\  (-l < y \le l)$ with two point singularities  put at $y=0$ and $l$.
Let us start with 
\begin{wrapfigure}[17]{l}{6.6cm}
\begin{center}
	\includegraphics[scale=0.8]{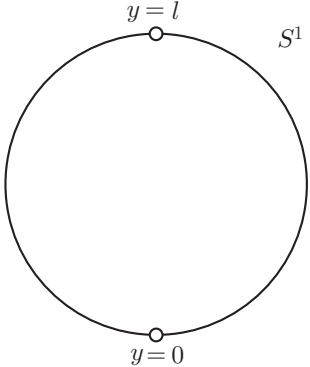}
\end{center}
	\caption{a circle $S^1(-l < y \le l)$ with two point singularities at $y=0$ and $l$.}
	\label{fig:1}
\end{wrapfigure} 
the following Hamiltonian :
 \begin{equation}
 	H=-\frac{d^2}{dy^2}+ \left(W'(y)\right)^2 - {\cal P}W''(y), 
\end{equation}
where a prime denotes the derivative with respect  to $y$, and ${\cal P}$ is the parity transformation defined by 
\begin{equation}
	{\cal P}: \varphi(y) \to {\cal P} \varphi (y) =\varphi(-y).
\end{equation}
The Hamiltonian forms the $N=2$ superalgebra \cite{Witten}
\begin{equation}
	\{ Q_a , Q_b \} =2H \delta_{ab}, \quad a,b=1,2
\end{equation}
with the supercharges \cite{N=2SUSY,extendedSUSY,Tsutsui}
\begin{eqnarray}
		Q_1&=&-{\cal P} \frac{d}{dy} +W'(y), \\
		Q_2 &=&i \frac{d}{dy} -i {\cal P} W'(y),
\end{eqnarray}
provided  that $W'(y)$ is an odd function of $y$, i.e. 
\begin{equation}
	W'(-y)=-W'(y).
\end{equation}
We should stress that to solve the eigenvalue problem 
\begin{equation}
	H\varphi_n(y)=m_n^2 \varphi_n (y)\label{6},
\end{equation}
the setting given above is insufficient.
Since we put singularities at $y=0$ and $l$, there are no reasons that wavefunctions as well as $W'(y)$ are smooth at the singularities, so that we should allow $\varphi_n(y)$ and $W'(y)$ to have discontinuities there. 
Therefore, to complete the setting of our model, we need to impose BC's at the singularities.
For BC's to be consistent with supersymmetry, we require $H$ and $Q_a (a=1,2)$ to be hermitian 
with respect to the inner product
\begin{equation}
	\langle f | g \rangle \equiv  \int_{-l_+}^{0_-} dy f^{*}(y) g(y)+\int_{0_+}^{l_-} dy f^*(y)g(y),
\end{equation}
where $0_{\pm} \equiv 0 \pm \varepsilon$ and $\pm l_{\mp} \equiv \pm l \mp \varepsilon$ with an infinitesimal positive constant $\varepsilon$.
Possible BC's that make the Hamiltonian hermitian are parameterized, in general, by the group $U(4)$.\footnote{In the case of a point singularity, allowed BC's are shown to be specified by $U(2)$ \cite{pointsingularity}.}
The further requirement that the supercharges are hermitian puts severe restrictions on allowed BC's.
In fact, there are only four types of BC's,  \cite{N=2SUSY,extended SUSY}\footnote{If we allow $\varphi (0_{\pm})$ and $\varphi'(0_{\pm}$) to be connected with $\varphi(\pm l_{\mp})$ and $\varphi'(\pm l_{\mp})$, we have a two parameter family of BC's \cite{extended SUSY}.} which are found to be consistent with the parity transformation, given by 
\begin{eqnarray}
	{\rm Type \ I\ }  \mbox{BC's}&:&\varphi'_{+}(0_+)+W'(0_+)\varphi_+(0_+)=0=\varphi_- (0_+), \nonumber \\
	&&\varphi'_+ (l_-)+W'(l_-)\varphi_+ (l_-)=0=\varphi_- (l_-),\label{8}\\
	{\rm Type \ II\ } \mbox{BC's}&:&\varphi'_{+}(0_+)+W'(0_+)\varphi_+(0_+)=0=\varphi_- (0_+), \nonumber \\
	&&\varphi_+ (l_-)=0=\varphi'_- (l_-)-W'(l_-)\varphi_-(l_-)\label{9},
\end{eqnarray}
where $\varphi_+(y) \left( \varphi_-(y)\right)$ denotes a parity-even (odd) function of $y$.
The other two BC's can be obtained by the replacement of $\varphi_+\leftrightarrow \varphi_-$ in eqs.$(\ref{8})$ and $(\ref{9})$.
It should be noted that the type I BC's reduce to the ordinary BC's for smooth parity eigenfunctions when $W'(0_+)=W'(l_-)=0$; otherwise, some of  the functions $\varphi_{\pm}(y)$ and $\varphi'_{\pm}(y)$ will have discontinuities at $y=0$ and $l$. 
Since the parity operator ${\cal P}$ commutes (anticommutes) with $H$ ($Q_a$),\footnote{This implies that ${\cal P}$ can be regarded as the \lq\lq fermion" number operator $(-1)^F$\cite{Witten}.} the energy spectrum is found to be doubly-degenerate between $\varphi_{+,n}$ and $\varphi_{-,n}$ (except for zero energy states).
For appropriately normalized eigenfunctions, we will have the relations 
\begin{equation}
	Q_1 \varphi_{\pm,n}(y)=m_n \varphi_{\mp,n}(y). \label{10}
\end{equation}
Similar relations hold for $Q_2$. 
The analysis of zero energy solutions needs some care. 
The fact that the functions $\varphi_{\pm,0}(y) \propto \exp{\{\mp W(y)\}}$, which are normalizable, satisfy $H\varphi_{\pm, 0}(y)=0$ does not necessarily imply the existence of zero energy solutions.
In fact, only $\varphi_{+,0}(y)\  (\varphi_{-,0}(y))$ is a zero energy solution for the model with the type I BC's (the type I BC's with the replacement of $\varphi_+ \leftrightarrow \varphi_-$), and there are  no other zero energy solutions at all.
This is because other solutions do not match with the desired BC's, so that they have to be eliminated from the spectrum.

For later use, it is convenient to rewrite the equations $(\ref{6})$ in a different basis such that
\begin{equation}
	-\frac{1}{\Delta_{\pm}(y)} \frac{d}{dy} \left(
		\Delta_{\pm} (y) \frac{d}{dy}\chi_{\pm,n}(y) \right)
	=m_n^2 \chi_{\pm,n}(y), \label{11}
\end{equation}
where
\begin{eqnarray}
	\chi_{\pm,n} (y)&=&e^{\pm W(y)} \varphi_{\pm,n}(y), \label{12}\\
	\Delta_{\pm}(y)&=& e^{\mp 2W(y)}.
\end{eqnarray} 
In this basis, the BC's $(\ref{8})$ and $(\ref{9})$ reduce to 
\begin{eqnarray}
	{\rm Type \ I \ }\mbox{BC's}&:& \chi'_+ (0_+)=0=\chi_-(0_+) ,\nonumber \\
	&&\chi'_+(l_-)=0=\chi _- (l_-),
\end{eqnarray}
\begin{eqnarray}
	{\rm Type \ II \ }\mbox{BC's}&:& \chi'_+ (0_+)=0=\chi_-(0_+) ,\nonumber \\
	&&\chi_+(l_-)=0=\chi' _- (l_-).
\end{eqnarray}
It follows that $\chi_{\pm}(y)$ simply obey either the Dirichlet or the Neumann BC at each singular point.
The relations $(\ref{10})$ are replaced by
\begin{equation}
	\chi'_{\pm, n}(y)=\pm m_n \Delta_{\mp} (y) \chi_{\mp, n}(y).
\end{equation}

To clarify characteristics of eigenvalues $m_n^2$ in eq.$(\ref{6})$ or $(\ref{11})$, let us examine a simple but still nontrivial example of the superpotential, $W(y)=c|y|$, in some detail.
Then, the eigenvalues and the associated eigenfunctions are found to be 
\begin{eqnarray}
	&&{\rm Type \ I \ model}: \nonumber \\
	&&\quad  0) \ m_0^2=0 \nonumber \\
	&& \qquad \varphi^{{\rm type\,  I}}_{+,0}(y) =\sqrt{\frac{c}{1-e^{-2cl}}}
		e^{-c |y|}, \\
	&&\quad  n) \ m_n^2=c^2+\left(\frac{n \pi}{l}\right)^2 \quad (n=1,2,3, \cdots)\nonumber \\
	&& \qquad \varphi^{{\rm type \, I}}_{+,n}(y) =\frac{1}{\sqrt{l}}
		\sin{\left(\frac{n \pi}{l}|y| -\theta_n\right)},\\
	&& \qquad \varphi^{{\rm type\,  I}}_{-,n}(y) =\frac{1}{\sqrt{l}}
		\sin{\left(\frac{n\pi}{l}y\right)}, \\
	&&\qquad {\rm with} 
	\  \tan{\theta_n}=\frac{n \pi}{c l}. \nonumber \\
	&&{\rm Type \ II \ model}: \nonumber \\
	&&\quad  0) \ m_0^2=c^2-K^2 \ {\rm with }\ K=c \tanh{(Kl)}  \\
	&& \qquad \varphi^{{\rm type \, II}}_{+,0}(y) =
		\left\{
			\begin{array}{ll}
				\sqrt{\frac{2K}{\sinh{(2K l)}-2K l}}
					\sinh{K(|y|-l)} 
	\qquad \ \ &{\rm for}\  cl>1,  \\
	 			\sqrt{\frac{3}{2l^3}}
					(|y|-l) 
	\qquad \qquad \qquad \qquad \qquad &{\rm for}\ cl=1,  \\
	 			non \qquad \qquad \qquad \qquad \qquad \qquad \qquad &{\rm for}\ cl<1,
			\end{array}
		\right.\\	
	&& \qquad \varphi^{{\rm type \, II}}_{-,0}(y) =
		\left\{
			\begin{array}{ll}
				\sqrt{\frac{2K}{\sinh{(2K l)}-2K l}} \sinh{Ky}  \qquad \qquad \quad \  &{\rm for}\ cl>1,  \\
	 			\sqrt{\frac{3}{2l^3}} y \qquad \qquad \qquad \qquad \qquad \qquad \quad &{\rm for}\ cl=1,  \\
	 			non \qquad \qquad \qquad \qquad \qquad \qquad \quad \quad & {\rm for}\ cl<1,
			\end{array}
		\right.\\	
	&&\quad  n) \ m_n^2=c^2+k_n^2 \quad (n=1,2,3, \cdots)\  {\rm with} \ k_n=c \tan{k_n l}\nonumber \\
	&& \qquad \varphi^{{\rm type\, II}}_{+,n}(y) =\sqrt{\frac{2k_n}{2k_n l-\sin{(2k_n l)}}}
		\sin{k_n (|y|-l)}, \\
	&& \qquad \varphi^{{\rm type \, II}}_{-,n}(y) =\sqrt{\frac{2k_n}{2k_n l-\sin{(2k_n l)}}}
		\sin{k_ny}.
\end{eqnarray}
The important observations of the above results are summarized as follows:
For the type I model, a zero energy solution $\varphi^{{\rm type \, I}}_{+,0}$ appears as a localized state around a singular point, and the higher modes 
\  (KK modes) are doubly-degenerate between $\varphi^{{\rm type \, I}}_{\pm,n}$ with eigenvalues $m_n^2=c^2 +\left(\frac{n\pi}{l}\right)^2$ for $n=1,2,3,\cdots$.
On the other hand, for the type II model the lowest eigenvalue for $cl \gg 1$ is non-vanishing but
\begin{equation}
	m_0^2 \sim 4c^2 e^{-2cl}. \label{mass_square}
\end{equation}
Thus, the eigenvalue is exponentially suppressed by the factor $e^{-2cl}$.
This turns out to be a key ingredient of our gauge symmetry breaking with a large mass hierarchy, as we will see in the next section. 
The higher modes have eigenvalues $c^2+k_n^2$, so that the lowest eigenvalue is much lighter than the others. 
Thus, the type II model is a realization of the idea given by Shaposhnikov and Tinyakov.
It should be noticed that for $c<0$ there is no state with the mass squared $(\ref{mass_square})$ and thus that all masses are heavier than $|c|$. 
In constructing phenomenological models, the lowest modes, which are massless for the type I BC's and are massive but much lighter than the others for the type II BC's, will survive at low energies.

\section{A model with a large mass hierarchy}

In order to demonstrate how Higgsless gauge symmetry breaking occurs with a large mass hierarchy, let us consider an $SU(5)$ gauge theory with a single extra dimension compactified on the orbifold $S^1/Z_2$, where $S^1$ is a circle with $-l < y \le l$ and $Z_2$ is defined by $y \to -y$.
\begin{eqnarray}	
S&=&\int d^4 x \int_{S^1/Z_2} dy \ \Delta(y) \left\{ -\frac{1}{4} F^a_{MN} F^{aMN}\right\} \nonumber \\
&=& \int d^4 x \int_{S^1/Z_2} dy \ \Delta(y) \left\{ -\frac{1}{4}F^{a}_{\mu \nu} F^{a\mu \nu}
             -\frac{1}{2}F^{a}_{5\nu} F^{a5\nu} \right\},
\end{eqnarray}
where $F^a_{MN} =\partial _M A^a_N -\partial _N A^a_{M}+g_5 f^{abc}A^b_M A^c_N $, and the $f^{abc}$'s are the structure constants of the gauge group $SU(5)$.
The five-dimensional coordinates are denoted by $x^M=(x^{\mu},y)$, where four-dimensional Lorentz indices are symbolized by Greek letters, e.g. $\mu,\nu =0,1,2,3$.
The $\Delta(y)$ is some weight function depending on the fifth coordinate $y$.
We assume $\Delta(y)$ to be an even function with the form $\Delta(y)=e^{-2W(y)}$, and we will take $W(y)=M_G |y|$, as an illustrative example, later.
Although we will not discuss the origin of the weight function, it may arise, for instance, in a way from compactification on 3-branes with the metric ansatz $ds^2=\Delta(y)^2 \left( \eta_{\mu \nu} dx^{\mu}dx^{\nu} - dy^2 \right)$,
\footnote{We note that the metric $ds^2 =e^{-4M_G |y|} (\eta_{\mu \nu}dx^{\mu}dx^{\nu}-dy^2 )$ resembles the warped metric 
$ds_{RS}^2 =e^{-2k|y|} \eta_{\mu \nu} dx^{\mu} dx^{\nu} -dy^2$, \cite{R-S} but gives a quite different mass spectrum, as we will see below.} 
or a dilaton background.\cite{dilaton}

The orbifold projection on the gauge fields is given by 
\begin{eqnarray}
	&&A_{\mu}(x,-y)= P A_{\mu}(x,y) P^{-1}, \\
	&&A_5(x,-y)=-P A_5(x,y)P^{-1},
\end{eqnarray}
where $A_M=A^a_M T^a $ ($T^a$'s are the generators of $SU(5)$), and $P$ is a $5\times 5$ matrix defined by $P=diag{(1,1,1,-1,-1)}$.
The above orbifold projection implies that the gauge fields $A^a_{\mu}(x,y)$ belonging to the $SU(3)\times SU(2) \times U(1)$ gauge symmetry are even functions of $y$, while the remaining gauge fields $A^{\hat{a}}_{\mu}(x,y)$ are odd functions of $y$.
On the other hand, the scalar components of the gauge fields have opposite parity; $A^a_5(x,y)$ $(A^{\hat{a}}_5(x,y))$ are odd (even) functions of $y$.
Thus, one might conclude that the $SU(5)$ gauge symmetry is reduced to $SU(3)\times SU(2) \times U(1)$ .
This is, however, a hasty conclusion.

The above setting of the orbifold model is not enough to get the mass spectrum in four dimensions.
Most of the literature have tacitly assumed that field configurations are smooth at fixed points of orbifolds. 
It will, however, natural to allow field configurations to be discontinuous at fixed points since orbifolds are singular there.\footnote{Some attempts \cite{bagger} to generalize orbifold models with discontinuities at fixed points have been made.}\nopagebreak
\ Thus, we here impose nontrivial BC's discussed in the previous section at the fixed points $y=0$ and $l$.
 
We would like to have a model that exhibits a desired GUT symmetry breaking, i.e.
\begin{equation}
	SU(5) \stackrel{M_G}{\longrightarrow} SU(3)\times SU(2)\times U(1)\stackrel{M_W}{\longrightarrow} SU(3)\times U(1)_{em}, 
\end{equation}
where $M_G$ denotes a GUT scale on the order of $10^{16}$ GeV and $M_W$ a weak scale on the order of $10^2$ GeV.
Let $A^{SU(3)\times U(1)_{em}}_M$ be the abbreviation of the gauge fields belonging to the unbroken gauge symmetry $SU(3)\times U(1)_{em}$, and let $A^{SU(5)/SU(3)\times SU(2)\times U(1)}_M $ and $A^{SU(2)\times U(1)/U(1)_{em}}_M$ be the abbreviations of the gauge fields of the broken generators corresponding to $SU(5) \to SU(3) \times SU(2) \times U(1)$ and $SU(2)\times U(1) \to U(1)_{em}$, respectively.
The lowest modes of $A^{SU(3)\times U(1)_{em}}_{\mu}$ have to be massless, and those of $A^{SU(2)\times U(1)/U(1)_{em}}_{\mu}$, which correspond to the $W^{\pm}$ and $Z^0$ bosons in the Standard Model, should have masses on the order of $M_W$.
On the other hand, all the modes of $A^{SU(5)/SU(3)\times SU(2)\times U(1)}_{\mu}$ are expected to have masses on the order of $M_G$ or heavier.
It turns out that the above situation can be realized if the gauge fields of the unbroken symmetry $SU(3)\times U(1)_{em}$ obey the type I BC's and the gauge fields belonging to the broken generators obey the type II BC's.
The mode expansions of the gauge fields are given by \footnote{Since the orbifold $S^1/Z_2$ is given by modding out $S^1$ under the parity reflection, the normalization of the eigenfunctions should be multiplied by $\sqrt{2}$.}
\begin{eqnarray}
	A^{SU(3)\times U(1)_{em}}_{\mu}(x,y)&=&\sum_{n=0}^{\infty}A^{SU(3)\times U(1)_{em}}_{\mu,n}(x)\  \chi^{{\rm type \, I}}_{+,n}(y), \\
	A^{SU(2)\times U(1)/U(1)_{em}}_{\mu}(x,y)&=&\sum_{n=0}^{\infty}A^{SU(2)\times U(1)/U(1)_{em}}_{\mu, n}(x)\  \chi^{{\rm type\, II}}_{+,n}(y), \\
	A^{SU(5)/SU(3)\times SU(2)\times U(1)}_{\mu}(x,y)&=&\sum_{n=1}^{\infty}A^{SU(5)/SU(3)\times SU(2)\times U(1)}_{\mu, n}(x) \ \bar{\chi}^{{\rm type \, II}}_{-,n}(y), 
\end{eqnarray}
and 
\begin{eqnarray}
	A^{SU(3)\times U(1)_{em}}_{5}(x,y)&=&\sum_{n=1}^{\infty}A^{SU(3)\times U(1)_{em}}_{5, n}(x)\  e^{2W(y)}\chi^{{\rm type\, I}}_{-,n}(y), \\
	A^{SU(2)\times U(1)/U(1)_{em}}_{5}(x,y)&=&\sum_{n=0}^{\infty}A^{SU(2)\times U(1)/U(1)_{em}}_{5, n}(x)\  e^{2W(y)}\chi^{{\rm type\, II}}_{-,n}(y), \\
	A^{SU(5)/SU(3)\times SU(2)\times U(1)}_{5}(x,y)&=&\sum_{n=1}^{\infty}A^{SU(5)/SU(3)\times SU(2)\times U(1)}_{5, n}(x) \ e^{2W(y)}\bar{\chi}^{{\rm type \, II}}_{+,n}(y), \nonumber \\
\end{eqnarray}
where $\chi_{\pm,n}(y)$ are defined through eq.$(\ref{12})$, and $\bar{\chi}_{\pm,n}(y)$ are also defined through eq.$(\ref{12})$ with the replacement of $W(y) \to -W(y)$.
It follows that all the massive modes are degenerate between $A_{\mu,n}$ and $A_{5,n}$, and
each massive mode of $A_{5,n}$ may be regarded as a supersymmetric partner of the 
longitudinal mode of $A_{\mu, n}$.\footnote{A similar supersymmetric structure has been found
for a 5D gauge theory with a general warped metric. \cite{lim}
}
Since every $A_{5, n}$ mode is massive, 
there are no massless 4D scalars in our model.
\begin{table}
     \caption{$Z_2$ parity, BC's and mass spectrum: $k_n \ (\bar{k}_n)$ for $n=1,2,3,\cdots$ are the solutions to $k_n=M_G \tan{k_n l} \ (\bar{k}_{n}=-M_G \tan{\bar{k}_n l}$).}
     \label{table:1}
     \begin{center}
       \begin{tabular}{cccl} \hline \hline 
        4D fields  & $Z_2$ parity  & BC's & mass spectrum \\ \hline \\
            $A^{SU(3)\times U(1)_{em}}_{\mu,n}$     &  $+$   &  type I  & $\begin{array}{l}m_0=0 \\
            m_n=\sqrt{M_G^2+(\frac{n\pi}{l})^2}\end{array}$
            \\\\
            $A^{SU(2)\times U(1)/U(1)_{em}}_{\mu,n}$     & $ +$   &  type II  &$\begin{array}{l}m_0\sim 2 M_G e^{-M_G l}  \\      m_n=\sqrt{M_G^2+k_n^2}\end{array}$
            \\\\
            $A^{SU(5)/SU(3)\times SU(2)\times U(1)}_{\mu,n}$     & $ -$   &  type II  &$\begin{array}{l}m_n=\sqrt{M_G^2 +(\bar{k}_n)^2}\end{array}$
            \\ \hline \\
            $A^{SU(3)\times U(1)_{em}}_{5,n}$     &  $-$   &  type I  &$\begin{array}{l}m_n=\sqrt{M_G^2 +(\frac{n\pi}{l})^2}\end{array}$
             \\\\
              $A^{SU(2)\times U(1)/U(1)_{em}}_{5,n}$     &  $-$   &  type II  & $\begin{array}{l}m_0 \sim 2 M_G e^{-M_G l}\\
            m_n=\sqrt{M_G^2+k_n^2}\end{array}$
            \\\\
            $A^{SU(5)/SU(3)\times SU(2)\times U(1)}_{5,n}$     &  $+$   &  type II  &$\begin{array}{l}m_n=\sqrt{M_G^2 +(\bar{k}_n)^2} \end{array}$
            \\ \hline
       \end{tabular}
     \end{center}
\end{table}

If we take $\Delta(y)=e^{-2M_G|y|}$ with $M_G\sim 10^{16}$ GeV and $M_G l\sim 33 $, the low energy effective theory consists of $A^{SU(3)\times U(1)_{em}}_{\mu,0}(x)$ and $A^{SU(2)\times U(1)/U(1)_{em}}_{\mu,0}(x)$.
The gauge fields $A^{SU(3)\times U(1)_{em}}_{\mu,0}(x)$ remain massless, while $A^{SU(2)\times U(1)/U(1)_{em}}_{\mu,0}(x)$, which correspond to the $W^{\pm}$ and $Z^0$ bosons of the Standard Model, acquire the masses on the order of $2 M_Ge^{-M_G l} \sim 10^2$ GeV.
The other modes get masses on the order of $M_G$ and decouple at low energies, as desired.

\section{Summary and discussions}

In this Letter, we have proposed a mechanism of gauge symmetry breaking with a large mass hierarchy.
As a demonstration, we have considered a 5D $SU(5)$ gauge theory with a single extra dimension of $S^1/Z_2$.
The orbifold has singularities at fixed points, and thus we can impose nontrivial BC's that make field configurations discontinuous at fixed points.
We have shown that the quadratic part of the Lagrangian possesses an $N=2$ supersymmetric structure in a quantum mechanical point of view.
This is the origin of the appearance of massless modes and exponentially small mass scale.

An extension to orbifolds $S^1/(Z_2\times \cdots \times Z_2)$ will be straightforward.
One-dimensional quantum mechanics on $S^1$ with $2^N$ point singularities has been investigated in a supersymmetric point of view. \cite{extendedSUSY}
The analysis can equally apply for orbifold models on $S^1/(Z_2\times \cdots \times Z_2)$, as
was done in this Letter.

To construct a realistic model, we have to introduce fermions.
In our approach, fermions can live in the bulk.
This is because nontrivial BC's at fixed points can assure chiral structure of 4D fermions localized around some fixed points.
Since in our model low energy modes of gauge fields appear to be localized around fixed points, it is suggested that all the contents of the Standard Model except for the Higgs scalars are given by localized modes around fixed points.
We view our model given in this Letter as a step toward a realistic theory of GUT symmetry breaking without Higgs scalars.
It would be interesting to explore orbifold models along the line discussed here.

We finally comment  on a question of the unitarity violation in our model.
The Higgs scalar in the Standard Model is necessary to ensure the unitarity of massive gauge boson scattering at high energies.\cite{unitarity-Higgs}
Without it, the scattering amplitude for the longitudinal components of the massive $W$ and $Z$ bosons would grow with energy as $\sim E^2$.
It has been shown \cite{unitarity-orbifold,Murayama} that higher dimensional gauge theories maintain unitarity in the sense that the terms in the amplitude that would grow with energy as $E^4$ or $E^2$ cancel, and there massive KK modes play a role of the Higgs scalar to ensure the unitarity.
In our model, the tree-level unitarity will be, however, violated
at the TeV scale, just like the Standard Model without
a Higgs scalar, because the KK modes which could cancel
the terms of the amplitude growing with energy as $E^2$ are
too heavy $(\sim M_G)$ to have much influence before unitarity
is violated.
Thus, our model will become a strong coupling theory
and cease a reliable perturbative framework at energies
above the TeV scale.
To make reliable calculations based on our scenario,
additional low energy degrees of freedom will be required
in order to unitarize the amplitude.\footnote{For instance, a scalar field living at a boundary
may rescue the low energy unitarity. \cite{Murayama}}
\section*{Acknowledgements}
The authors wish to thank
Y. Hosotani,
Y. Kawamura,
C. S. Lim,
N. Maru,
M. Tachibana,
K. Takenaga,
S. Tanimura, 
I. Tsutsui 
for valuable discussions and useful comments. 
This work was supported in part by the Grant-in-Aid for
Scientific Research (No.15540277) by the Japanese Ministry
of Education, Science, Sports and Culture.


\end{document}